\documentclass[aps,prl,twocolumn,superscriptaddress]{revtex4}

\usepackage{graphicx}
\usepackage[german,english]{babel}
\newcommand{\imp}{\ensuremath{\text{imp}}}
\newcommand{\hc}{\ensuremath{\text{h.c.}}}

\begin{document}
\title{Monovalent impurities on graphene: midgap states and migration barriers}

\author{T. O. Wehling}
\email{twehling@physnet.uni-hamburg.de}
\affiliation{I. Institut f{\"u}r Theoretische Physik, Universit{\"a}t Hamburg, Jungiusstra{\ss}e 9, D-20355 Hamburg, Germany}
\author{M. I. Katsnelson}
\affiliation{Institute for Molecules and Materials, Radboud
University of Nijmegen, Heijendaalseweg 135, 6525 AJ Nijmegen, The
Netherlands}
\author{A. I. Lichtenstein}
\affiliation{I. Institut f{\"u}r Theoretische Physik, Universit{\"a}t Hamburg, Jungiusstra{\ss}e 9, D-20355 Hamburg, Germany}

\pacs{72.80.Rj; 73.20.Hb; 73.61.Wp}


\date{\today}

\begin{abstract}
Monovalent impurities on graphene can be divided into ionically
and covalently bond impurities. The covalent impurities cause
universal midgap states as the carbon atom next to the impurity is
effectively decoupled from the graphene $\pi$ bands. The
electronic structure of graphene suppresses migration of these
impurities and making the universal midgap very stable. This
effect is strongest for neutral covalently bond impurities. The
ionically bond impurities have migration barriers of typically
less than $0.1$\,eV. An asymmetry between anions and cations
regarding their adsorption sites and topology of their potential
energy landscape is predicted.
\end{abstract}
\maketitle

The discovery of graphene \cite{Novoselov_science2004} and its
remarkable electronic
properties \cite{reviewGK,reviewktsn,RMP_AHC2007} initiated great
research interest in this material. Particularly prospective for
applications is the extraordinarily high charge carrier mobility $\mu$
in graphene \cite{Novoselov_science2004,bolotin2008,DSBA08}. In
combination with a very high Fermi velocity $v \approx 10^6$ m/s
this makes micron mean free paths routinely achievable.

Away from the neutrality point, the conductivity of graphene is
weakly temperature dependent and approximately proportional to the
carrier concentration $n$ \cite{kostya2,kim}. The mechanism
limiting the electron mobility in graphene is still under debate.
Charged impurities are probably the simplest and thus the most
natural candidate \cite{NM06,Ando06,AHGS07}. However,
room-temperature experiments with gaseous adsorbates such as
NO$_2$ have showed only a weak dependence of $\mu$ on charged
impurity concentration \cite{Schedin07}. Furthermore, recent
experiments \cite{Mohiuddin09} did not find any significant
dependence of $\mu$ on immersing graphene devices in high-$\kappa$
media such as ethanol and water (dielectric constants $\kappa
\approx 25$ and $80$, respectively) but this disagrees with
another report \cite{Fuhrer_PRL08} in which a few monolayers of ice
increased $\mu$ in graphene by $\approx 30 \%$. Because of the
experimental controversy, alternative mechanisms such as
scattering on frozen ripples \cite{KG08} and resonant
impurities \cite{KN07,SPG07} were discussed. However, both these
mechanisms also have some weak points, rather theoretical than
experimental. There is still no real theory which would explain
why the ripple structure becomes quenched and thus almost
temperature independent. As for the resonant scatterers, in
general, the closeness of the impurity quasilocal states to the
neutrality point necessary to make this mechanism efficient looks
just accidental. To discuss this as the main scattering mechanism
one needs, at least, to clarify the mechanism which makes the
resonant scatterers typical for graphene. Thus, the physics of
charge carrier mobility in graphene, crucially important for most
of potential applications, is not clarified yet. Also, while
impurities appear as undesirable residua from the graphene
production process, chemical functionalization of graphene relies
on impurities for controlling its electronic properties as
demonstrated recently for hydrogenated graphene
(graphane) \cite{graphane09}.

For judging which impurities might determine electron scattering in graphene and for optimizing chemical functionalization, the mechanisms determining the impurity mobility need to be known. In this letter, we consider monovalent adsorbates and show that these can be divided into two separate groups regarding the bonding mechanism: ionically and covalently bond impurities. To this end, we present ab-initio calculations on H, Li, Na, K, Cs, F, Cl, Br, I, CH$_3$ and OH adsorbates on graphene. For these systems the electronic structure and migration barriers are analyzed. The covalently bond impurities cause a characteristic midgap state derived from the graphene electrons. This state turns out to be very stable, as graphene's conjugated $\pi$ bonds enhance the migration barriers of neutral covalently bond impurities.

For a first principles description of the graphene adsorbate systems we performed density functional calculations within the generalized gradient approximation (GGA)\cite{Perdew:PW91, PBE} on $4\times 4$ graphene supercells containing one impurity. The Vienna Ab Initio Simulation Package (VASP) \cite{Kresse:PP_VASP} with the projector augmented wave (PAW) \cite{Bloechl:PAW1994,Kresse:PAW_VASP} basis sets has been used for solving the resulting Kohn-Sham equations. In this way we obtained relaxed structures for the graphene adsorbate systems, total energies, and orbitally resolved local density of electronic states (LDOS).

\begin{figure}[ht]
\begin{minipage}{0.49\linewidth}
 \raggedright
 a) 
\includegraphics[width=.98\linewidth]{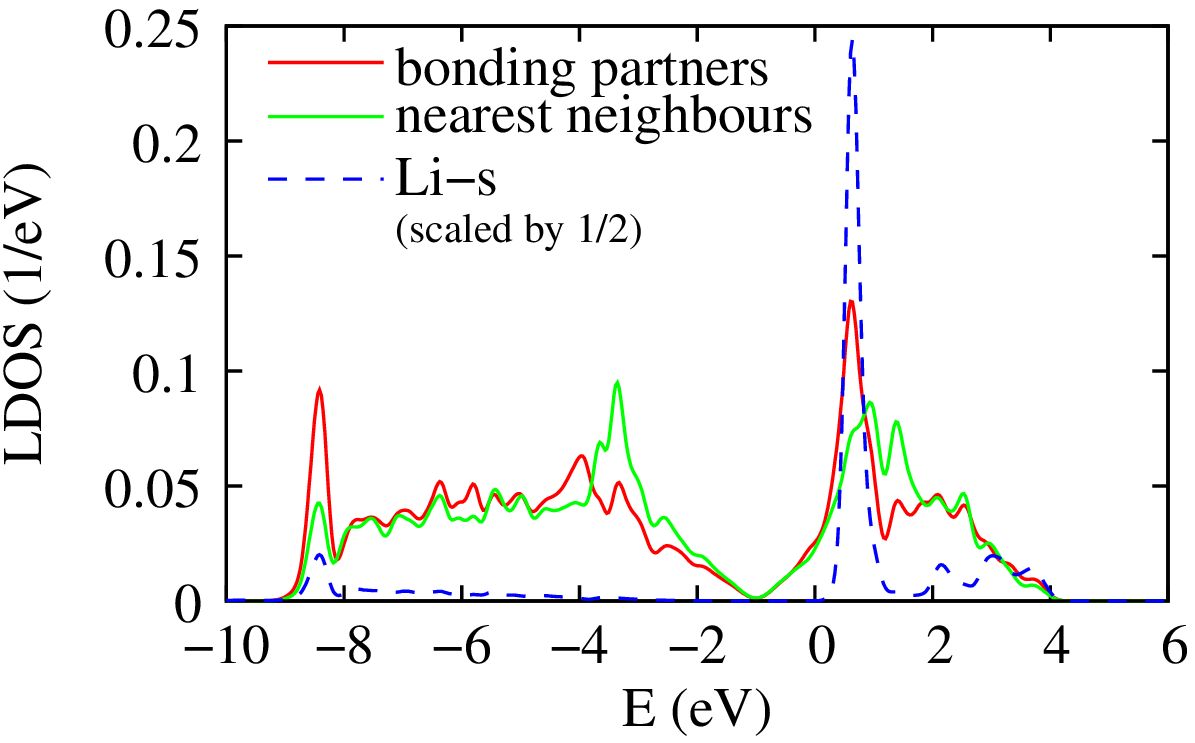}
\end{minipage}
\begin{minipage}{0.49\linewidth}
 \raggedright
  b) \includegraphics[width=.98\linewidth]{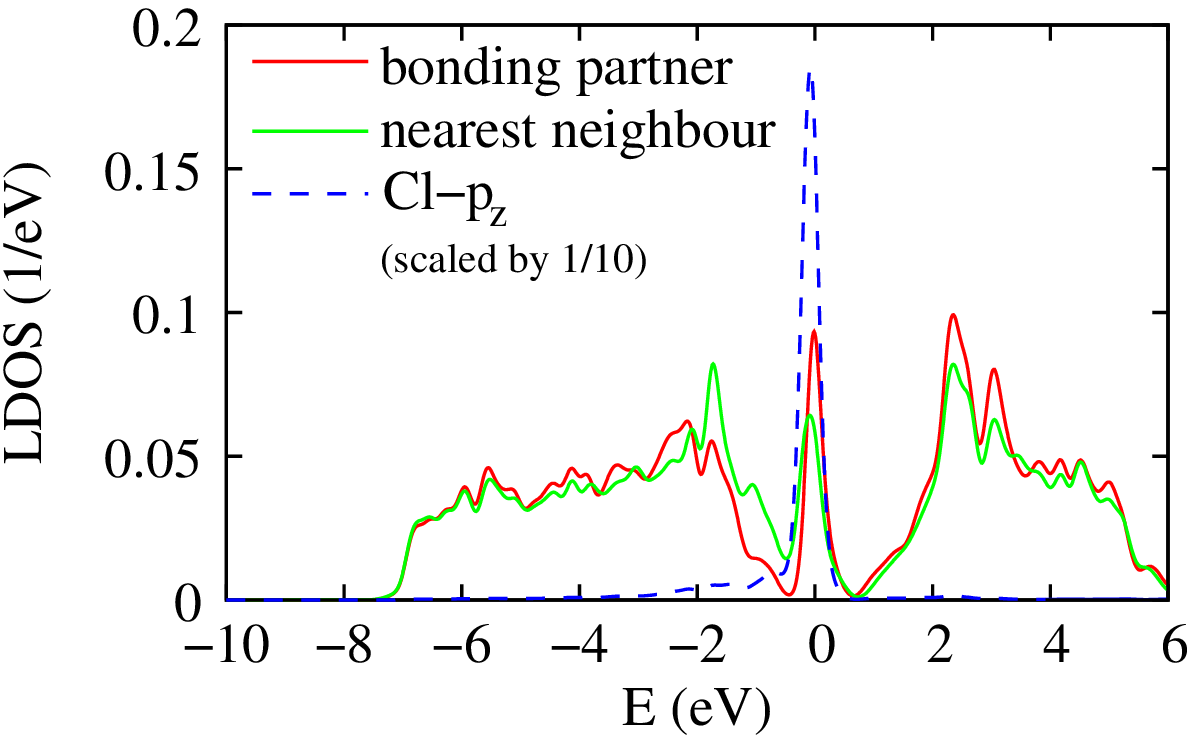}
\end{minipage}
 \begin{minipage}{0.49\linewidth}
 \raggedright
  c) \includegraphics[width=.98\linewidth]{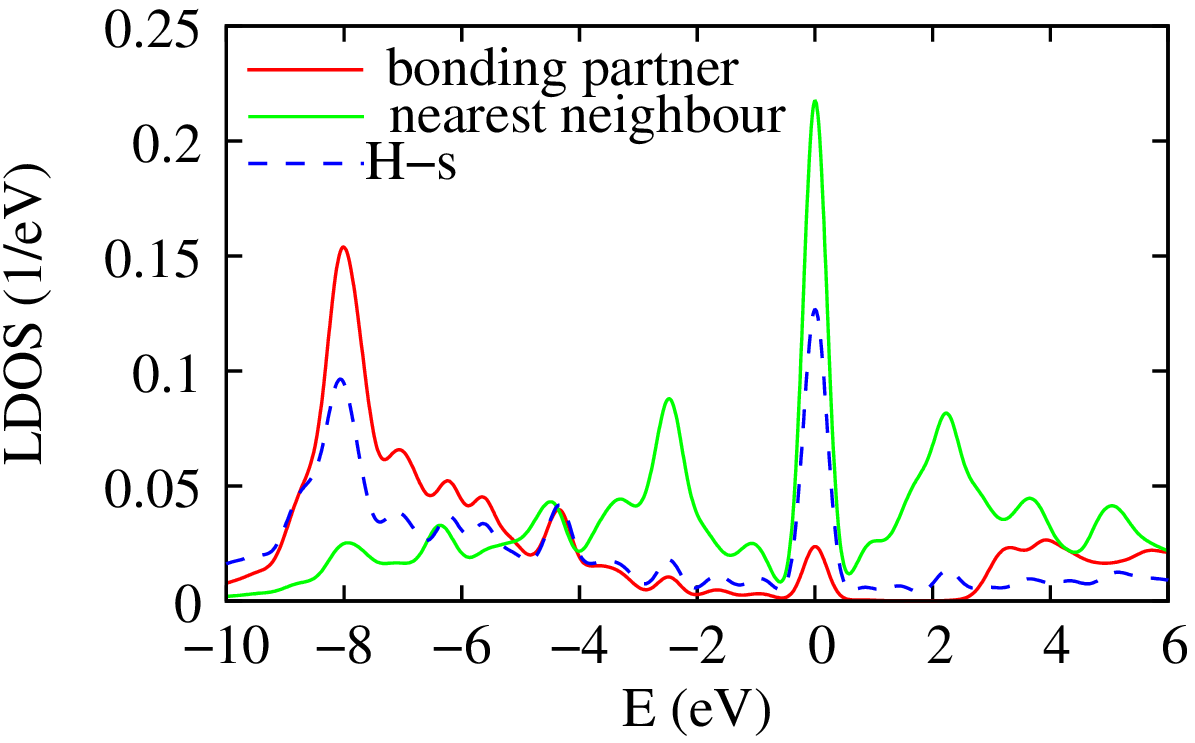}
\end{minipage}
\begin{minipage}{0.49\linewidth}
 \raggedright
  d) \includegraphics[width=.98\linewidth]{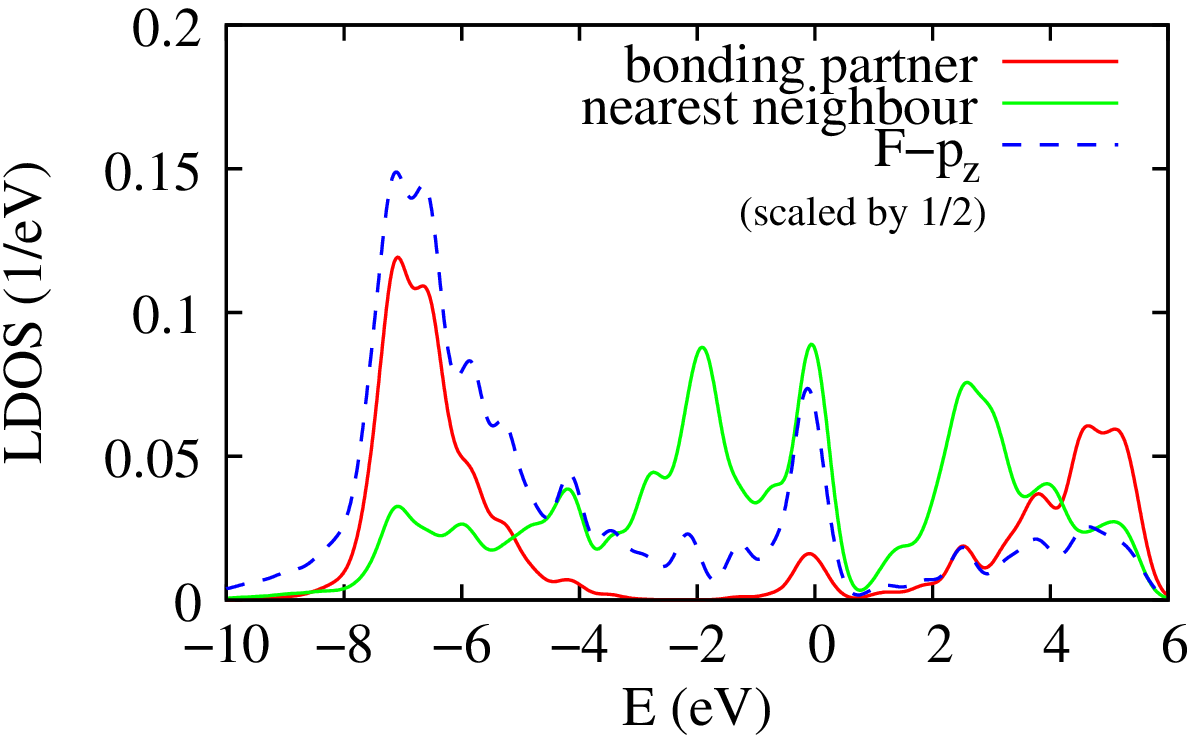}
\end{minipage}
\begin{minipage}{0.49\linewidth}
 \raggedright
  e) \includegraphics[width=.98\linewidth]{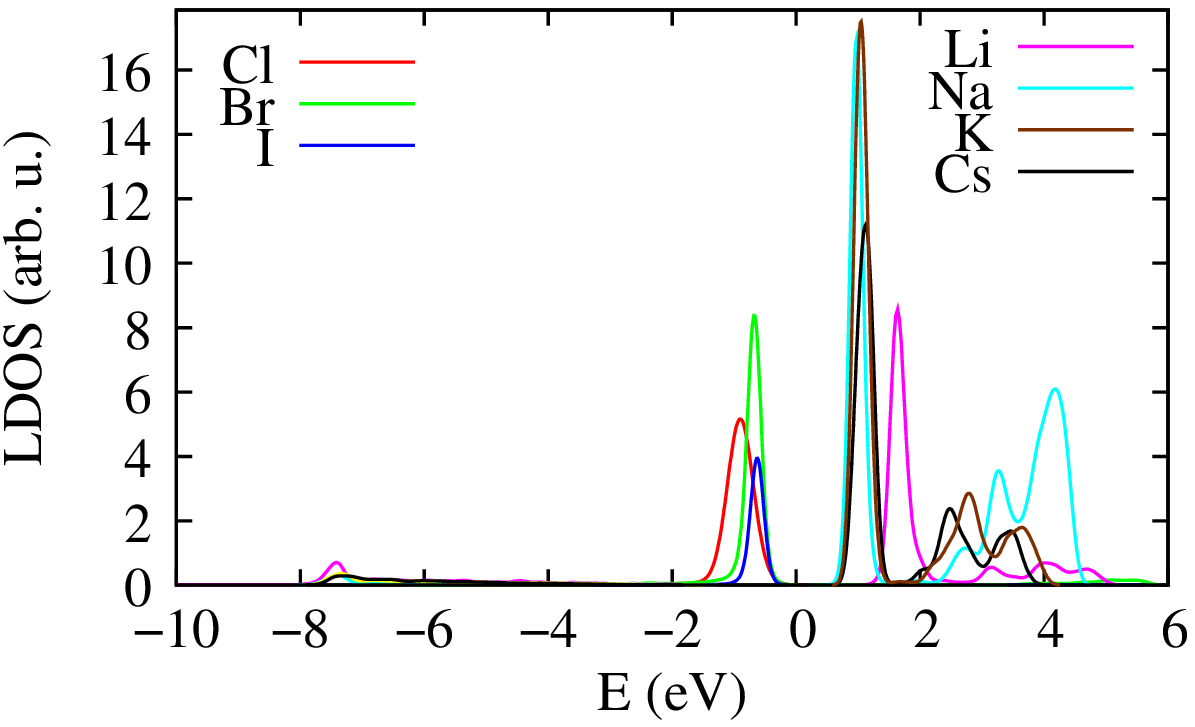}
\end{minipage}
\begin{minipage}{0.49\linewidth}
 \raggedright
  f) \includegraphics[width=.98\linewidth]{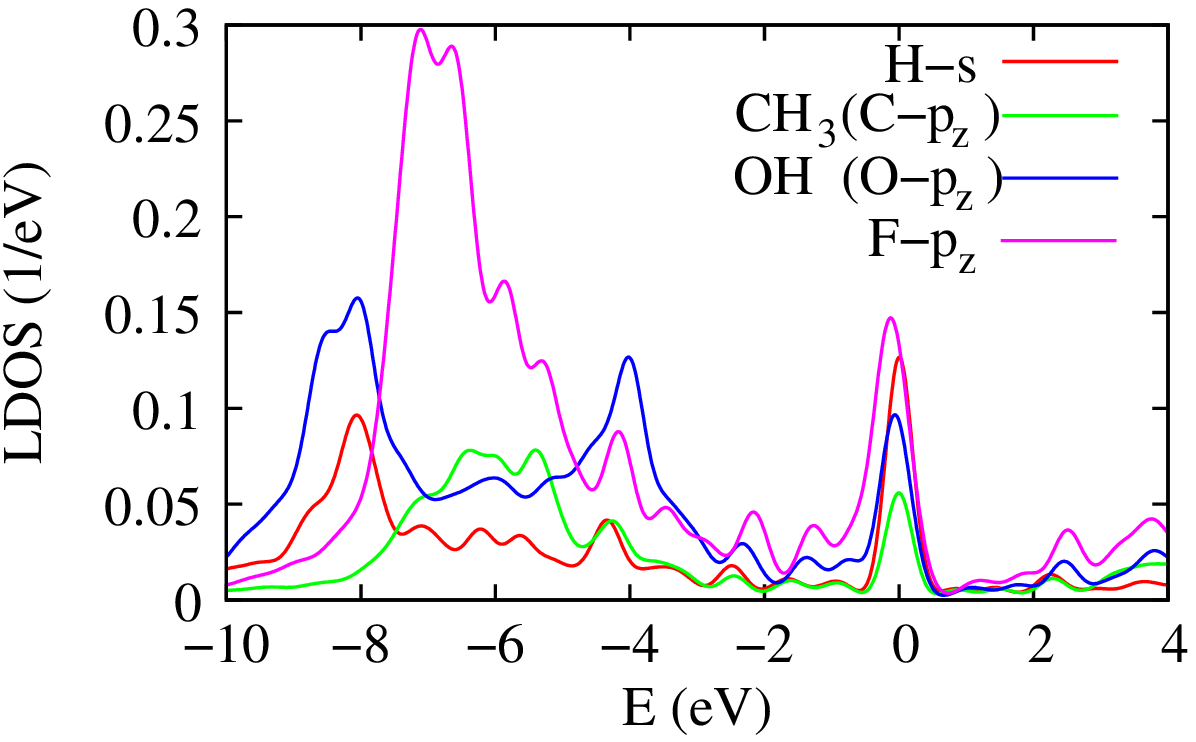}
\end{minipage}
 %
%
%
%
%
%
%

%
 %
%
\caption{\label{fig:DOS}(Color online) LDOS in different graphene adsorbate systems. (a-b) ionically bond impurities, (c-d) covalently bond impurities. a) Graphene + Li, b) Graphene + Cl, c) Graphene + H d) Graphene + F. For the impurity's bonding partner in graphene and its nearest neighbor the p$_z$ projected LDOS is shown. The valence electron LDOS at the impurity site for ionically bond impurities is depicted in (e) and for covalently bond impurities in (f). In (a-d) and (f) the Fermi level is at $E=0$; in (e) the Dirac point is at $E=0$.}
\end{figure}

The local electronic structure of graphene in the vicinity of adsorbates (Fig. \ref{fig:DOS}) can be grouped into two classes. The LDOS in the vicinity of adsorbates like Li or Cl exhibits a sharp resonance close to Fermi level which is almost entirely localized at the impurity. Besides this peak, the LDOS at the nearest neighbor and at the next nearest neighbor of the impurity exhibits the pseudogap characteristic for graphene.
This is qualitatively different for the second group of impurities (Fig. \ref{fig:DOS} c) and d)). H and F adatoms cause a midgap state characteristic for Dirac fermions: With the bonding partner of the impurity in sublattice A the impurity state is localized in sublattice B and at the impurity atom.

Every stable atomic configuration under investigation can be strictly grouped either into the class of strongly or weakly hybridized impurities, as can be seen from Fig. \ref{fig:DOS} e)-f): Ionically bond impurities give rise to a sharp acceptor (donor) level below (above) the Dirac point at $E_{\rm D}=0$. The LDOS of covalently bond impurities is much broader and exhibits characteristic resonances far below the Fermi level (between $-10$\,eV and $-4$\,eV) as well as a midgap state at the Fermi level.

\begin{figure}[ht]
\centering
\begin{minipage}{0.63\linewidth} \raggedright
  a) \includegraphics[width=.98\linewidth]{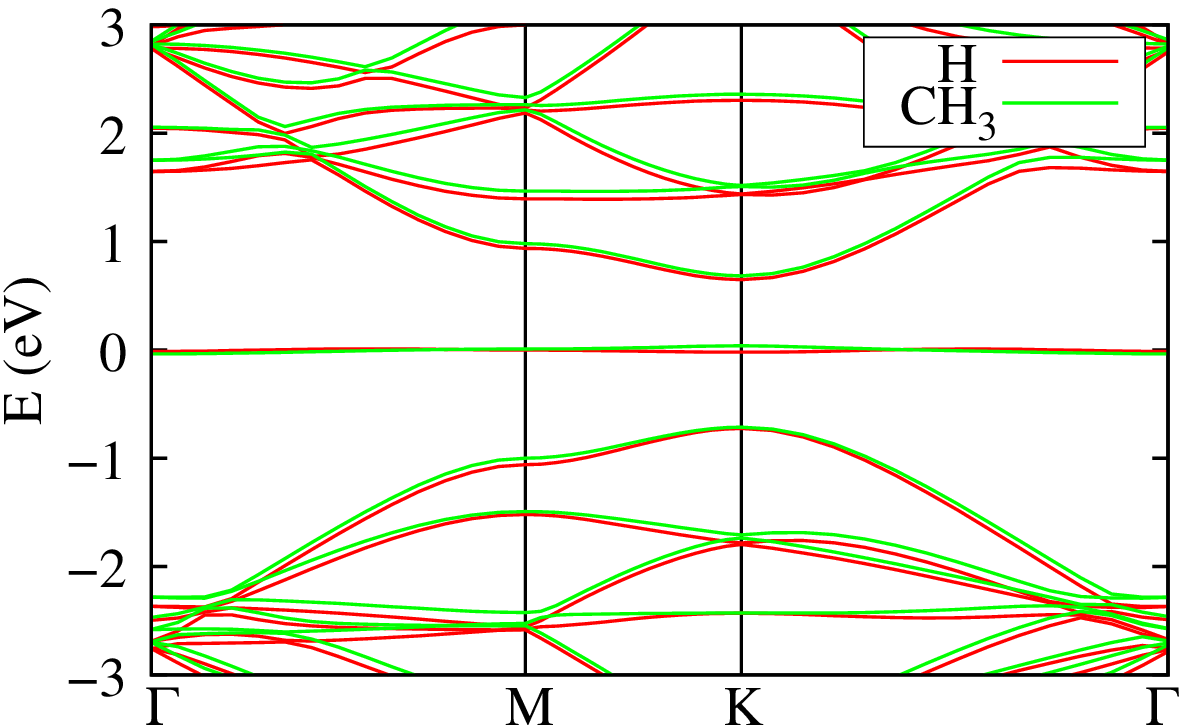}
\end{minipage}
\begin{minipage}{0.63\linewidth} \raggedright
  b) \includegraphics[width=.98\linewidth]{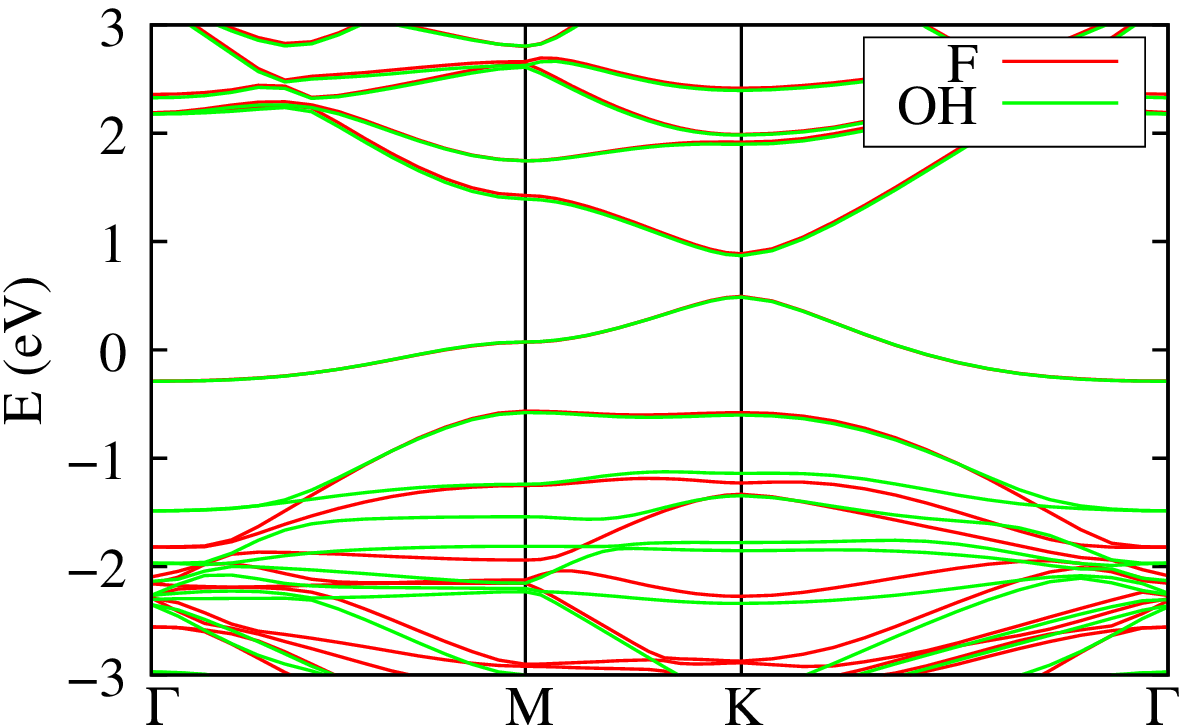}
\end{minipage}
\caption{\label{fig:bands}(Color online) Supercell band structure of a) H and CH$_3$ on graphene as well as b) F and OH bond covalently to graphene. Bands coincide close to the Fermi level ($E=0$) despite the different internal structure of the impurities.}
\end{figure}

As regards electron scattering this midgap state is mainly independent of the particular type of covalent impurity. The supercell band structures for H, CH$_3$, OH, and F covalently bond to graphene are shown in Fig. \ref{fig:bands}.
The band structure of graphene with adsorbed H and CH$_3$ as well as those for graphene with F and OH adsorbates coincide close to the Fermi level despite the different internal structure of the adsorbates. The coupling of the midgap state and the graphene bands can be quantified in an \textit{effective} impurity model $H=H_{\rm D}+H_{\imp}$, where the unperturbed graphene bands are described by $H_D=\sum_k\epsilon(k)d_k^\dagger d_k$ and the perturbation by
\[H_{\imp}=\epsilon_{\imp} c^\dagger c+\sum_k V_k c^\dagger d_k+\hc .\] Here, the index $k=(\vec {k},\nu)$ denotes crystal momentum $\vec {k}$ and band number $\nu=\pm$. $\epsilon(k)$ is the unperturbed graphene dispersion. The effective impurity is characterized by its energy $\epsilon_{\imp}$ and its hybridization $V_k$ with the graphene bands. In a supercell calculation at the backfolded Dirac points $\vec{k}=K^\pm$, this model simplifies to
\begin{equation}
\label{eqn:Himp}
 H=\left(\begin{array}{ccc}\epsilon_{\imp} & A & B \\ A^* & 0 & 0 \\ B^* & 0 & 0\end{array}\right),
\end{equation}
where the zero block stems from the graphene bands at the Dirac point and $A$ ($B$) are the  components of $V_k$ in the two different sublattices A (B). This allows to derive the coupling strengths $|V|=\sqrt{|A|^2+|B|^2}$ and $\epsilon_{\imp}$ from the DFT energies of the three bands closest to the Fermi level in the supercell calculation as shown in table \ref{tab:imp_param}.

\begin{table}
\caption{\label{tab:imp_param} Impurity energy and hybridization for the effective impurity model of the midgap state for different impurities. All impurities are placed on top of a C atom, which is at the total energy minimum for the covalently bond impurities and the anions Cl / Br but not for the cations Li and Na.}
\begin{ruledtabular}
\begin{tabular}{l|cc||l|cc}
& $\epsilon_{\imp}$ (eV) & $|V|$ (eV)&  & $\epsilon_{\imp}$ (eV) & $|V|$ (eV)\\
\hline
H & -0.03 & 1.38 & Li & 1.17 & 0.22\\
CH$_3$ & -0.11 & 1.40 & Na & 0.93 & 0.13\\
OH & -0.70 & 1.30 & Cl & -0.79 & 0.43\\
F & -0.67 & 1.30 & Br & -0.73 & 0.19

 \end{tabular}
\end{ruledtabular}
 \end{table}
The coupling $|V|$ is mainly independent of the internal structure of the covalently bond impurities.
Hence, this midgap state appears as a universal feature of all monovalent impurities which are strongly bond to one of graphene's carbon atoms.

Bonding H-atoms to graphene and related electron scattering has been analyzed in \cite{boukhvalov:08,casolo-2008,robinson08}. For the bonding partner of H, the $\pi$ bond to its nearest carbon neighbors is broken and a $\sigma$ bond with the H ad-atom is formed. The carbon bonding partner of H atoms has been found to be decoupled from the graphene $\pi$-electron system and the resulting local imbalance between the number of atoms belong to each of the two sublattices causes a midgap state. The band structures from Fig. \ref{fig:bands} and the coupling constants from Table \ref{tab:imp_param} show that same mechanism is effective for all monovalent covalently bond impurities on graphene.

In the following we show, that the creation of this midgap state by an impurity covalently bond to one carbon atom enhances migration barriers for covalently bond monovalent impurities in graphene. A comparison to ionically bond impurities is given.
To find migration barriers for ionically bond impurities it is sufficient to perform structural relaxations with the impurities in three different high symmetry adsorption sites: on top of a C-atom (t-site), in the middle of a hexagon (h-site) and above the middle of a nearest neighbor C-C bond (b-site). The covalent impurities cause strong distortion of the nearby bonds and require the minimum energy paths to be calculated using the nudged elastic band method \cite{NEB}.

In agreement with \cite{Cohen08}, we find the energy minimum for the alkali cations at the h-sites and barriers as shown in table \ref{tab:imp_barrier}. The barriers decrease with cation size and are all (except for the special case of Li) below 0.1 eV.
\begin{table}
\caption{\label{tab:imp_barrier} Migration barriers $\Delta E$ and minimum energy site for ionically and covalently bond impurities.}
\begin{ruledtabular}
\begin{tabular}{l|cc||l|cc||l|cc}
& site & $\Delta E$ (eV) & & site & $\Delta E$ (eV) & & site & $\Delta E$ (eV)\\
\hline
H & t & 1.01 & Li & h & 0.31 & Cl & t,b & $<0.005$\\
CH$_3$ & t & 0.63 & Na & h & 0.09 & Br & t,b & $<0.005$\\
OH & t & 0.53 & K & h & 0.06 & I & t,b & $<0.005$\\
F & t & 0.29 & Cs & h & 0.04 &
 \end{tabular}
\end{ruledtabular}
 \end{table}
The potential energy landscape for the cations consists of dips in the center of the hexagons bordered by a hexagonal net of banks. Within this net spanned by the nearest neighbor carbon bonds, the variation of potential energy is by a factor of more than 5 smaller than between the h-site and the t/b-sites.

This landscape is reversed for the anions: Having their energy minima on the net and maxima in the center of the hexagons, the anions can freely move on the graphene sheets. 
The fact that the height of the impurity above the sheet is always minimized in center of the hexagon, would result in the energy minimum being in the center of the hexagon for all ionically bond impurities if atomic scale inhomogeneities in the screening charge of the impurities were negligible. The anions preferring the t- and b-sites over the h-sites shows that inhomogeneities in the screening charge corrugate the potential energy landscape of the ions on the order some 10 meV.

While this can be quite significant in the case of purely ionically bond impurities our calculations show that the potential energy landscape for covalently bond impurities is by an order of magnitude more corrugated. We find migration barriers between $0.29$\,eV for F and $1.01$\,eV for H. Notably, F is the impurity with the highest absolute binding energy ($1.99$\,eV) of all impurities considered, here, but it has the smallest migration barrier within the group of covalent impurities. For F and OH we find the saddle point energy of the transition path significantly below the desorption energy, which is $0.91$\,eV for OH. This is in strong contrast to H and CH$_3$: For H the energy of the saddle point state is  only 4 meV below the desorption barrier and moving a CH$_3$-group from one carbon atom to its nearest neighbor requires even overcoming the desorption barrier of $0.63$\,eV. No saddle point configuration with the CH$_3$-group in the middle of two neighboring C-atoms except for the CH$_3$ being fully desorbed from the graphene sheet could be found.

\begin{figure}[ht]
\centering
\begin{minipage}{0.65\linewidth}
 \raggedright  a) \centering \includegraphics[width=.98\linewidth]{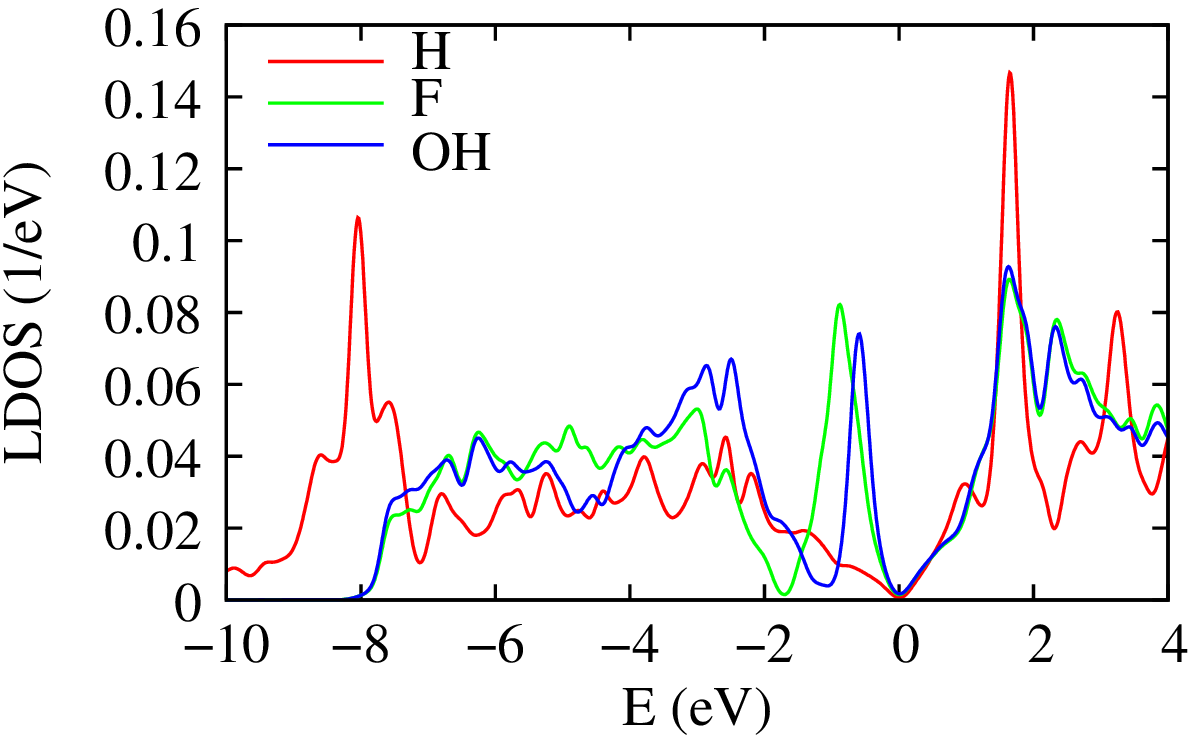}
\end{minipage}
\begin{minipage}{0.49\linewidth}\raggedright
  b) \includegraphics[width=.99\linewidth]{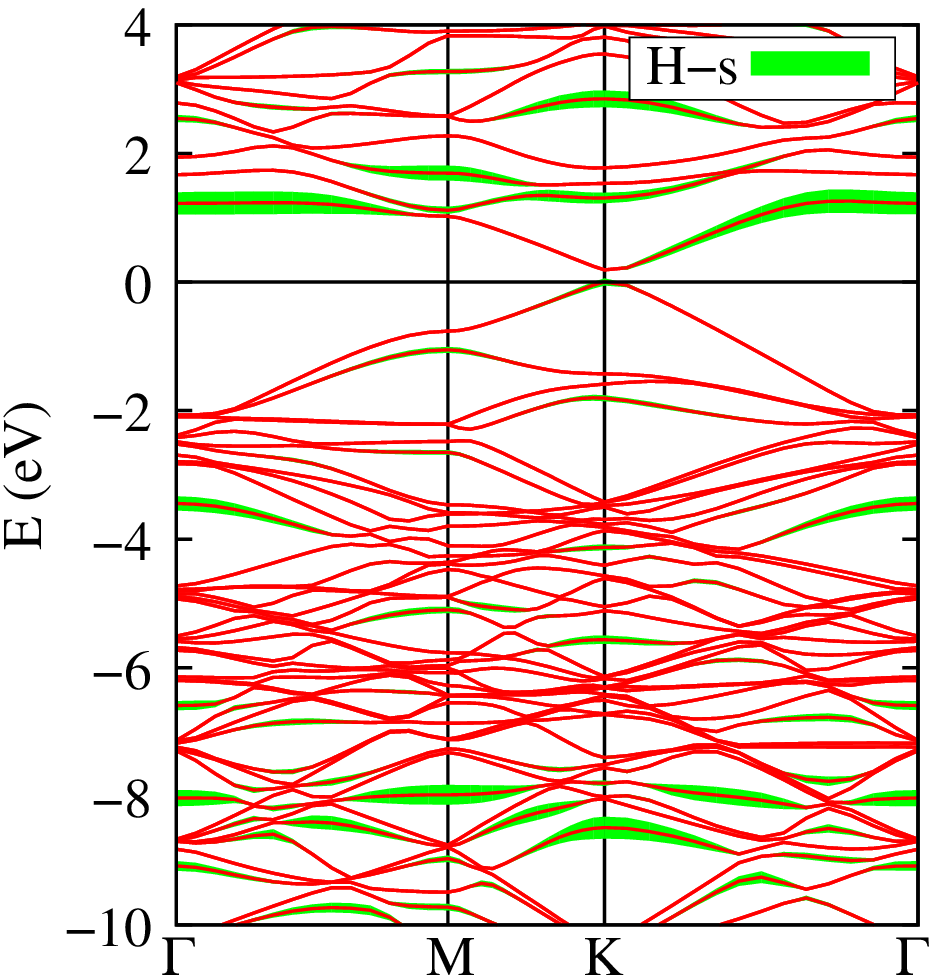}
\end{minipage}
\begin{minipage}{0.49\linewidth}\raggedright
  c) \includegraphics[width=.99\linewidth]{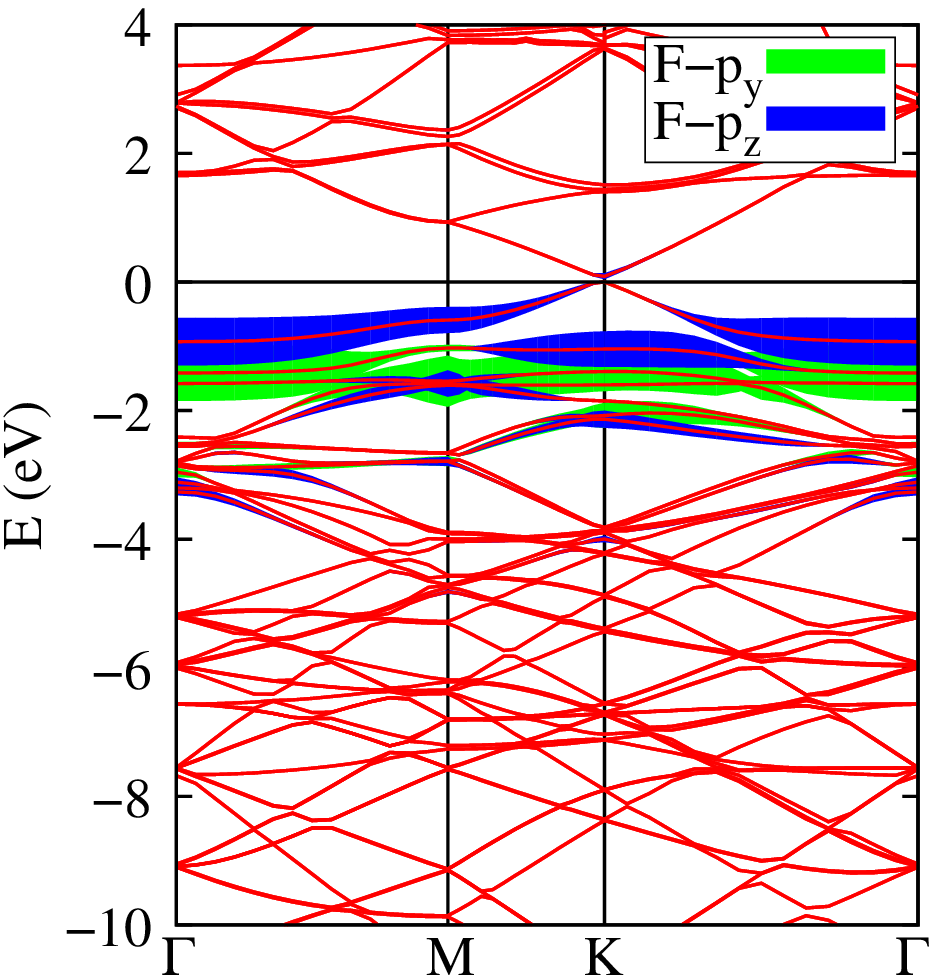}
\end{minipage}
\caption{\label{fig:saddle}(Color online) Electronic structure in the transition state. a) $p_z$-LDOS at a carbon atom next to the impurity. Lower part: Band structures of graphene supercell with b) H and c) F impurities in transition state configuration. Contributions from the impurity atoms are marked as fatbands.}
\end{figure}

For F, OH, and H the nature of chemical binding in the saddle point configuration can be understood from the supercell electronic properties shown in Fig. \ref{fig:saddle}. The LDOS at the carbon atoms next to the F and OH impurities is very similar to the LDOS in the vicinity of \textit{ionically} bond impurities like Cl or Br (see Fig. \ref{fig:DOS}b). This is in contrast to the case of H, where in addition to a resonance at $2$\,eV, the LDOS at the carbon neighbor of the impurity is broadened and exhibits a peak at $-8$\,eV --- similar to all covalently bond impurities in their minimum energy configuration. The H-impurity causes a donor level and is at the same time covalently bond, as the supercell band structure with contributions from the H impurity marked as fat bands further illustrates. There are contributions from the H $s$-orbital over the energy range from -10eV up to +3eV, indicating strong hybridization of the impurity orbital with the graphene bands. This is very different from F in the saddle point configuration with its valence orbitals contributing significantly within an energy interval which is an order of magnitude smaller. In the transition state F and OH are ionically bond to graphene.

The high barrier for H suggests that the formation of a strong
covalent bond in the transition state is highly unfavorable. The
origin of this effect can be understood from the model
Hamiltonian, Eqn. \ref{eqn:Himp}: With the impurity on top of the
bridge site, sublattice symmetry is preserved: $A=B$ in Eqn.
(\ref{eqn:Himp}). The symmetric combination of the two C-$p_z$
orbitals adjacent to the impurity
$\phi_{+}=\frac{1}{\sqrt{2}}(0,1,1)$ will couple to the impurity
$\phi_{\imp}=(1,0,0)$. The antisymmetric combination
$\phi_-=\frac{1}{\sqrt{2}}(0,1,-1)$ is decoupled and forms the
analog of the midgap state occuring for the impurity on top of a
carbon atom: In the latter case, with the impurity's bonding
partner in sublattice A, one obtains $B=0$ and finite $A$ in Eqn.
(\ref{eqn:Himp}). Thus, $\phi_0=(0,0,1)$ is decoupled from the
impurity in the stable configuration. $\phi_0$ is non-bonding and
is therefore at the energy of the Dirac point.
$\phi_-=\frac{1}{\sqrt{2}}(0,1,-1)$, however, is an antibonding
combination of neighboring C-p$_z$ orbitals. The ab-initio
calculations show that the resonances derived from this state are
more than $1$\,eV above the Dirac point, unoccupied and not available
for screening the additional positive charge brought by the
H-impurity. Thus, the creation of a local charge is enforced by
graphene's electronic structure for the impurity in a b-site
saddle point configuration. As a consequence, a strong tendency to
ionic bonding with graphene \textit{decreases} migration barriers,
while migration preferably neutral covalently bond impurities is
suppressed. 

This tendency explains experimental findings of charged impurities moving almost freely on graphene \cite{Schedin07,Caragiu05} and experiments suggesting considerable migration barriers for H adsorbates \cite{MeyerCrommie08,graphane09}. The fact that clusterization of impurities
on graphene strongly suppresses their contribution to the
resistivity \cite{cluster09} makes covalently bond impurities
one natural candidate to main source of scattering limiting the
electron mobility in graphene. It is essential that, as demonstrated here,
these impurities frequently have quasilocal peaks nearby the
neutrality point --- not accidentally but enforced by symmetry.

The authors thank A. Geim for inspiring discussions. Support from SFB 668 (Germany), FOM (The Netherlands) as well as computer time from HLRN are acknowledged.
\bibliography{water_s}

\end{document}